\documentclass[aps,prl,twocolumn]{revtex4}
\usepackage{graphicx}

\begin{document}

\title{Signal of Bose condensation in an optical lattice at finite temperature}
\author{W. Yi\footnote{\textit{Current Address: Institut f\"{u}r Theoretische Physik,
Universit\"{a}t Innsbruck, Technikerstrasse 25, A-6020 Innsbruck, Austria}}, G.-D. Lin, and L.-M. Duan}
\affiliation {FOCUS center and MCTP, Department of Physics,
University of Michigan, Ann Arbor, MI 48109}

\begin{abstract}
We discuss the experimental signal for the Bose condensation of cold
atoms in an optical lattice at finite temperature. Instead of using
the visibility of the interference pattern via the time-of-flight
imaging, we show that the momentum space density profile in the
first Brillouin zone, in particular its bimodal distribution,
provides an unambiguous signal for the Bose condensation. We confirm
this point with detailed calculations of the change in the atomic
momentum distribution across the condensation phase transition,
taking into account both the global trapping potential and the
atomic interaction effects.

\end{abstract}

\maketitle

A recent work raised an interesting question on how to
unambiguously confirm the Bose condensation associated with a
superfluid phase at finite temperature for cold atoms in an
optical lattice \cite{1,2}. The signal of Bose condensation in an
optical lattice is typically connected with the interference peaks
in the time-of-flight images. It is shown in \cite{1} that for a
thermal lattice gas above the condensation temperature, one could
also see the interference peaks with a pretty good visibility
\cite{3}. This thermal visibility goes down considerably if one
takes into account a number of practical effects, as shown in the
recent work \cite{4}. On the other hand, however, the thermal
visibility above the transition temperature is not negligible, and
its value depends on the details of a number of system parameters.
There is no explicit critical value yet for the interference
visibility to separate the condensation region from the thermal
region.

In this work, we suggest to use the momentum space density profile
in the first Brillouin zone (measured from the time-of-flight
imaging) as an alternative method to characterize the Bose
condensate in an optical lattice. In particular, similar to the
free-space case, the bimodal distribution of the atomic momentum
distribution should provide an unambiguous signal for the Bose
condensation. To show the practicality of this method, we calculate
the finite-temperature atomic momentum distribution in a
three-dimensional (3D) optical lattice, confirming a distinctive
change in the distribution across the condensation phase transition.
There are several complexities in this calculation: first of all,
one needs to take into account the inhomogeneity of the lattice due
to the global harmonic trap, which tends to broaden the momentum
distribution of the gas, in particular for the condensate part.
Secondly, the atomic interaction leads to two competing effects: it
broadens the real-space distribution and thus sharpens the momentum
space density profile in a trap; while at the same time, it also
leads to the broadening of the momentum distribution during the
time-of-flight process. In our calculation, we take all these
effects into account.

To illustrate the basic character of the experimental signal, first
we calculate the finite temperature momentum distribution for free
bosons in an optical lattice with a global harmonic trap, in which
case exact solutions are possible after introducing some tricks.
With the Feshbach resonance technique, one can also directly test
the predictions in this case by turning off the atomic interaction.
Then we use the Hatree-Fock-Bogliubov-Popov (HFBP) approximation
\cite{5,6,7} to deal with the interaction between the atoms. The
HFBP method is expected to be a good approximation for weakly
interacting bosons except for a small region across the transition
from the superfluid phase to the normal phase \cite{5}. For cold
atoms in a lattice with a global harmonic trap, such a region
corresponds to only a slim layer, which has negligible influence on
the integrated momentum distribution. So we expect the HFBP
approximation should be a reasonably good method for the calculation
of the atomic momentum distribution before the Mott transition shows
up in the trap. We note that the HFBP approximation has been used
before for cold atoms, with the focus on the homogeneous case
\cite{6} or the one-dimensional system \cite{7}.

For free bosonic atoms in an inhomogeneous optical lattice with a
global harmonic trap, the Hamiltonian has the form $H=\int
d\mathbf{r}\Psi ^{\dag }( \mathbf{r})[-(\hbar ^{2}/2m)\partial
_{\mathbf{r}}^{2}+V_{op}+V(\mathbf{r} )]\Psi (\mathbf{r})$, where
$m$ is the atomic mass, $V_{op}=V_{0} \sum_{i=1,2,3}\sin ^{2}(\pi
r_{i}/d)$ is the optical lattice potential with $ d$ as the lattice
spacing, and $V(\mathbf{r})=\frac{1}{2}m\omega ^{2}\mathbf{ r}^{2}$
is the global harmonic trapping potential. To diagonalize the
Hamiltonian, we expand the atomic field operator $\Psi (\mathbf{r})$
as $ \Psi
(\mathbf{r})=\sum_{\mathbf{R}}w(\mathbf{r}-\mathbf{R})a_{\mathbf{R}}$,
where the operator $a_{\mathbf{R}}$ ($a_{\mathbf{R}}$) annihilates
(creates) a particle on site $R$, $w(\mathbf{r}-\mathbf{R})$ is the
Wannier function on the site $\mathbf{R}$ associated with the
lattice potential $V_{op}$, and the summation runs over all the
lattice sites. A Fourier transform of this expansion gives
\begin{equation}
\Psi (\mathbf{k})=w(\mathbf{k})a_{\mathbf{k}},
\end{equation}
where $\Psi (\mathbf{k}),w(\mathbf{k}),a_{\mathbf{k}}$ denote respectively
the Fourier components of $\Psi (\mathbf{r}),w(\mathbf{r}),a_{\mathbf{R}}$
in the momentum space.

Under typical experimental situations, the global harmonic trap $V(\mathbf{r}%
)$ varies slowly on individual lattice sites. In that case, under
the expansion above, the Hamiltonian is recast into the form
\begin{equation}
H=\sum_{\mathbf{k}}E_{\mathbf{k}}a_{\mathbf{k}}^{\dag }a_{\mathbf{k}}+\sum_{%
\mathbf{R}}V(\mathbf{R})a_{\mathbf{R}}^{\dag }a_{\mathbf{R}},
\end{equation}
where the summation over $\mathbf{k}$ runs over the first Brillouin zone. We
have assumed here that the system temperature is well below the band gap, so
the atoms only occupy the lowest band, where $E_{\mathbf{k}}$ is well
approximated by $E_{\mathbf{k}}=-2t\sum_{i=1,2,3}\cos (k_{i}d)$. The
parameter $t$ denotes the tunneling rate over the neighboring sites, with $%
t\approx \left( 3.5/\sqrt{\pi }\right) V_{0}^{3/4}e^{-2\sqrt{V_{0}}}$ for an
optical lattice \cite{8}, where both $t$ and $V_{0}$ are in the unit of the
recoil energy $E_{R}\equiv \hbar ^{2}\mathbf{\pi }^{2}/2md^{2}$.

The Hamiltonian (2) in principle can be directly diagonalized
numerically. However, the calculation of the finite temperature
momentum distribution requires determination of all the eigenstates
of the Hamiltonian, which is very time consuming for a
three-dimensional lattice with many sites. Here, we adopt an
approach which allows easy calculation of any finite temperature
momentum distributions \cite{add1,add2,add3}. Note that the indices
$\mathbf{R}$ and $\mathbf{k}$ in Eq. (2) are reminiscent of the
coordinate and the momentum variables in quantum mechanics. So we
can write down a first quantization Hamiltonian corresponding to Eq.
(2) in the momentum space, where $\mathbf{R}$ is replaced by the
momentum derivative $\partial _{\mathbf{k}}$. The resulting
Hamiltonian takes the form
\begin{equation}
H_{eff}=-\frac{1}{2}m\omega ^{2}\partial _{\mathbf{k}}^{2}+E_{\mathbf{k}}.
\end{equation}
This Hamiltonian describes free particles with effective mass $m^{\ast
}=\hbar ^{2}/(m\omega ^{2})$ trapped in an effective potential $E_{\mathbf{k}%
}$ with periodic boundary condition (the period is given by the
reciprocal lattice vector $\mathbf{K}$). The quasi-momentum
distribution $\left\langle a_{\mathbf{k}}^{\dag
}a_{\mathbf{k}}\right\rangle $ is then given by the
square of the eigenstate wavefunctions $\left| \phi _{\mathbf{n}}\left( \mathbf{k}%
\right) \right| ^{2}$ of $H_{eff}$, averaged over all the eigen-levels $%
\mathbf{n}$ with a Bose distribution factor $g(\epsilon _{\mathbf{n}%
})=1/\exp \left[ \left( \epsilon _{\mathbf{n}}-\mu \right)
/T-1\right] $ at finite temperature $T$, where $\epsilon
_{\mathbf{n}}$ is the corresponding eigenenergy and $\mu $ is the
chemical potential fixed by the total number of atoms. From Eq. (1),
we then obtain the atomic momentum distribution
\begin{eqnarray}
n(\mathbf{k}) &=&\left\langle \Psi ^{\dagger }(\mathbf{k})\Psi (\mathbf{k}%
)\right\rangle =\left| w(\mathbf{k})\right| ^{2}\left\langle a_{\mathbf{k}%
}^{\dag }a_{\mathbf{k}}\right\rangle  \nonumber \\
&=&\left| w(\mathbf{k})\right| ^{2}\sum_{\mathbf{n}}g(\epsilon _{\mathbf{n}%
})\left| \phi _{\mathbf{n}}\left( \mathbf{k}\right) \right| ^{2}
\end{eqnarray}
The signal from the time-of-flight imaging corresponds to the column
integrated momentum distribution $n_{\perp }(k_{x},k_{y})=\int n(\mathbf{k}%
)dk_{z}$.

Through Eq. (4), we have calculated the momentum distribution of a
free Bose gas in an optical lattice with a global harmonic trap at
different temperatures around the condensation transition. The
parameters are chosen to be close to those of a typical experiment
for $^{87}$Rb, with the lattice barrier $V_{0}=10E_{R}$ and the
total particle number $N\simeq 10^{5}$. The results for the
integrated momentum distribution $n_{\perp }(k_{x},k_{y})$ are
shown in Fig. 1 under different trap frequencies. In a weak
harmonic trap, there are distinctive interference peaks even for a
thermal gas above the transition temperature (Fig. 1(a)), which
agrees with the result of Ref. \cite {1} for a homogeneous optical
lattice that corresponds to the zero-trap limit. These peaks are
caused by the short range thermal correlations between different
sites. The correlation function of the thermal gas decays
exponentially with distance $x$ by the form $e^{-x/L}$, where the
characteristic length $L$ is estimated to be $\sim 1.1d$ for Fig.
1(a). As the trap frequency increases, the thermal interference
peaks become less visible and eventually disappear (Fig. 1(c)),
which agrees with the recent calculation of the visibility in Ref.
\cite{4}. In the case of a weak global trap, it gets difficult to
use the visibility to distinguish the condensation phase
transition as this parameter has a pretty high value for both the
condensed and the non-condensed phases near the critical
temperature \cite{1,4}. However, if one directly compares the
interference patterns in Figs. 1 (a,c) and (b,d) across the
condensation transition, the difference in the interference peaks
is still pretty clear: in particular, for the case with a
condensate component, the interference peaks become much sharper.
This reminds us to look at the momentum space density profile in
the first Brillouin zone, which, compared with the single
parameter of visibility, should be able to give more detailed
information about the system and its phase transition.

\begin{figure}[tbp]
\includegraphics{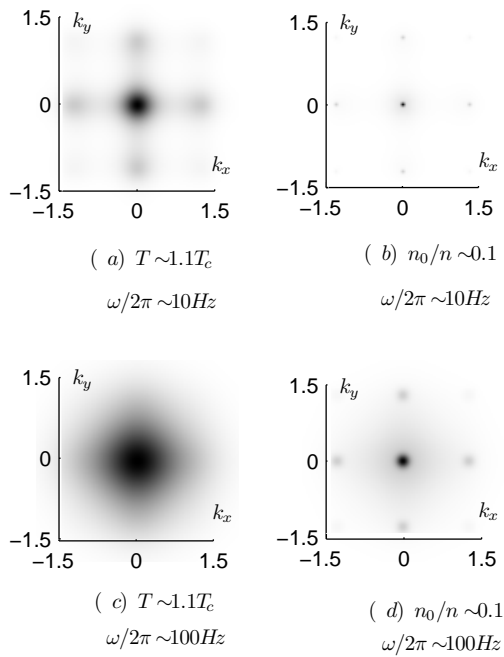}
\caption[Fig.1 ]{The intensity plot for the column integrated
momentum distribution near the first Brillouin zone with different trapping frequency $\protect%
\omega/2\pi$ and temperature $T$. Figs. (a) and (c) are for the
normal phase, and (b) and (d) are for the superfluid phase with the
corresponding parameters: (b) $T/T_c\sim 0.7$; (d) $T/T_c\sim 0.94$.
The total particle number is fixed to be $N_t=10^5$ for all the
cases, and the lattice barrier $V_0=10E_R$. The unit for the axes is
$2\pi/d$.}
\end{figure}

With this in mind, we plot the column integrated momentum
distribution along one of the axes passing through the center of
the first Brillouin zone. The results are plotted in Fig. 2 (a)
(c) (e) for different condensate fractions. It is clear that while
the density profile is a thermal distribution for a normal gas, a
bimodal structure appears when a non-zero condensate fraction
exists in the lattice. Based on this observation, we propose to
use the interference peaks as well as the bimodal structure in the
first Brillouin zone as an unambiguous signature for detection of
the Bose condensation in an optical lattice.

The calculation above is useful for an illustration of the general
qualitative features of the system, but we neglect the interaction
between the atoms in the lattice. In a more realistic model, such
interactions need to be taken into account, as they can greatly
modify the time-of-flight images. Firstly, the repulsive interaction
will tend to broaden the spatial distribution of the atoms and hence
narrow the momentum distribution in the trap. Secondly, in the
process of the time-of-flight expansion, the remnant atomic
interaction transforms the interaction energy into the kinetic
energy, which shows up in the final imaging process and broadens the
momentum distribution. In the presence of interaction, the system
can not be solved exactly. In the following, we calculate the
momentum distribution of an interacting Bose gas in an optical
lattice with a global harmonic potential trap using the
Hatree-Fock-Bogliubov-Popov (HFBP) approximation \cite{5}.

To calculate the momentum space density profile for an atomic gas in
an inhomogeneous optical lattice, we combine the HFBP method with
the local density approximation (LDA). Under the LDA, a local region
of the harmonic trap is treated as a homogeneous lattice system. We
first look at the Hamiltonian with inter-atomic interaction for a
homogeneous lattice gas:
\begin{equation}
H=-t\sum_{\langle \mathbf{i,j}\rangle }a_{\mathbf{i}}^{\dag }a_{\mathbf{j}}+%
\frac{U}{2}\sum_{\mathbf{i}}a_{\mathbf{i}}^{\dag }a_{\mathbf{i}}^{\dag }a_{%
\mathbf{i}}a_{\mathbf{i}}-\mu \sum_{\mathbf{i}}a_{\mathbf{i}}^{\dag }a_{%
\mathbf{i}}
\end{equation}
where $t$ is the tunneling rate defined before, $\mu $ is the local
chemical potential, $U=U_{bg}\int |w(\mathbf{r})|^{4}d\mathbf{r}$ is
the on-site interaction rate with an approximate form of
$U/E_R\approx 3.05(V_{0}/E_{R})^{0.85}(a_s/d)$ ($U_{bg}$ is related
to the s-wave scattering length by $U_{bg}=4\pi \hbar ^{2}a_{s}/m$,
and $a_s=5.45nm$ for $^{87}$Rb \cite{1,3}). We then transform all
the field operators to the momentum space, and write the momentum component $%
a_{\mathbf{k}}$ as $a_{\mathbf{k}}=v\delta _{\mathbf{k0}}+\delta a_{\mathbf{k%
}}$ with the standard HFBP approach, where $v\equiv \langle a_{\mathbf{0}%
}^{\dag }\rangle \equiv \left\langle a_{\mathbf{0}}\right\rangle $
represents the condensate fraction, and $\delta a_{\mathbf{k}}$ and
$\delta a_{\mathbf{k}}^{\dag }$ represent the excitations above the
mean field. We
keep the Hamiltonian to the quadratic order of the operators $\delta a_{%
\mathbf{k}}$ and $\delta a_{\mathbf{k}}^{\dagger }$, and then
diagonalize it to get the thermal potential $\Omega =-T\ln tr\left(
e^{-H/T}\right) $. The stationary condition $\partial \Omega
/\partial v=0$ gives the saddle point equation for the chemical
potential:
\begin{equation}
\mu =E_{0}-Un_{0}+2Un,
\end{equation}
where $E_{0}=-6t$ for a three-dimensional system, $n_{0}$ is the
per-site density of the condensate fraction, and $n$ is the total
number of particles per site. With the thermal potential, it is easy
to derive the number equation and the atomic momentum distribution.
For the non-condensate part with $\mathbf{k}\neq 0$, the per-site
quasi-momentum density distribution $\left\langle
a_{\mathbf{k}}^{\dag }a_{\mathbf{k}}\right\rangle $ is given by
\begin{equation}
\left\langle a_{\mathbf{k}}^{\dag }a_{\mathbf{k}}\right\rangle =\frac{E_{%
\mathbf{k}}-E_{0}+Un_{0}}{2\hbar \omega _{\mathbf{k}}}\coth \left( \frac{%
\hbar \omega _{\mathbf{k}}}{2T}\right) -\frac{1}{2},
\end{equation}
where $\hbar \omega _{\mathbf{k}}=\sqrt{(E_{\mathbf{k}%
}-E_{0})^{2}+2Un_{0}(E_{\mathbf{k}}-E_{0})}$ is the dispersion
relation for the Bogliubov excitations. With the LDA, the local
chemical potential $\mu ( \mathbf{r})$ at a displacement
$\mathbf{r}$ from the trap center is determined from $\mu (
\mathbf{r})=\mu _{0}-V(\mathbf{r})$, where $\mu _{0}$ is the
chemical potential at the trap center. We fix the total number
density per-site at the trap center, which, together with Eqs. (6-7)
and the number equation $n=n_0+\sum_{\mathbf{k}\neq0}\left\langle
a_{\mathbf{k}}^{\dag}a_{\mathbf{k}}\right\rangle$, allow us to
calculate the chemical potential $\mu_0$.  We may then determine
$\mu (\mathbf{r})$ and $\left\langle a_{\mathbf{k}}^{\dag
}a_{\mathbf{k}}\right\rangle $ at any trap location $\mathbf{r}$.
The overall non-condensate part of the atomic
momentum distribution is given by the integration of $\left\langle a_{%
\mathbf{k}}^{\dag }a_{\mathbf{k}}\right\rangle $ over the whole
harmonic trap, multiplied by the Wannier function $\left|
w(\mathbf{k})\right| ^{2}$ as shown in Eq. (4).

\begin{figure}[tbp]
\includegraphics{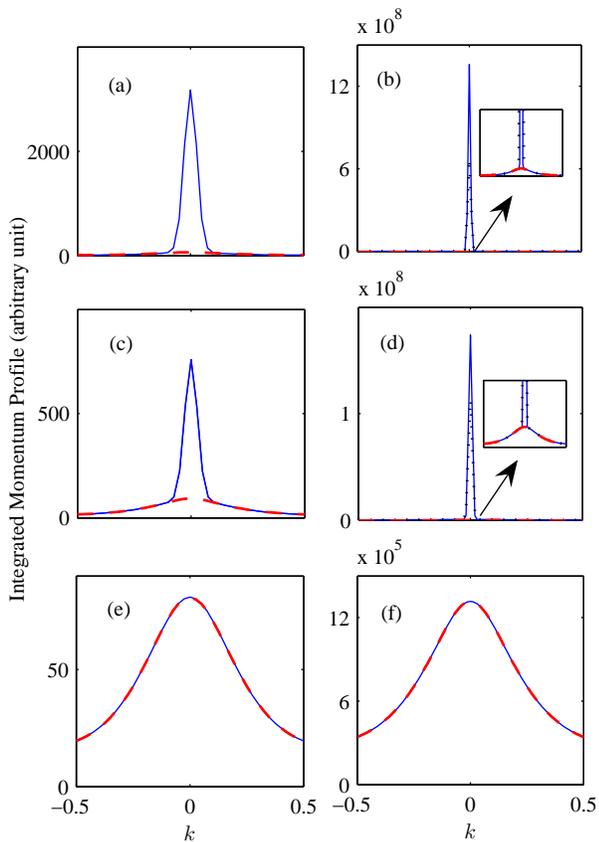}
\caption[Fig.2 ]{(Color online) Column integrated momentum density
profile (shown along the x-axis) for (a)(c)(e) non-interacting
lattice bosons and (b)(d)(f) interacting lattice bosons. The solid
curves are the total momentum density profile; the thick dashed
curves are the momentum density profile of the normal component; the
dotted curves are the momentum density profile taking into account
of the interaction broadening during the time-of-flight evolution.
(a) $N_{0}/N_{t}\sim 0.46$, $T/T_c\sim 0.75$ ($N_{0}$ is the total
number of atoms in the condensate and $N_{t}$ is the total number of
atoms in the trap); (b) $N_{0}/N_{t}\sim 0.52$, $T/T_c\sim 0.34$;
(c) $N_{0}/N_{t}\sim 0.10$, $T/T_c\sim 0.95$; (d) $N_{0}/N_{t}\sim
0.10 $, $T/T_c\sim 0.65$, inset: enlarged density profile below the
dotted line; (e) $T\sim 1.1T_{c} $; (f) $T\sim 1.1T_{c}$. For the
non-interacting calculations, the total particle number is fixed to
be $N_{t}=10^{5}$, and the trapping frequency is $\omega/2\pi=20Hz$;
for the calculations with interactions, the trapping potential is
$\omega/2\pi=20Hz$, the number density per-site at the center of the
trap is
fixed to be $n(0)=1$, with the total number of particles on the order of $%
N_{t}=10^{5}\sim 10^{6}$. The lattice barrier in all the cases is
$V_0=10E_R$. The unit for the x-axes is $2\pi/d$.}
\end{figure}

For the condensate part, the LDA will lead to an artificial $\delta $%
-function at zero momentum. To avoid this artifact, one needs to
consider explicitly the broadening of the condensate momentum
distribution by the harmonic trap. From the above LDA formalism, we
get the condensate fraction $ n_{0}(\mathbf{r})$ at any trap
location $\mathbf{r}$. The condensate wavefunction in the trap can
be well approximated by $\phi _{0}(\mathbf{r})=
\sqrt{n_{0}(\mathbf{r})}$ (which actually corresponds to the
solution of the Gross-Pitaevskii equation under the Thomas-Fermi
approximation \cite{9}). The condensate part of the atomic momentum
distribution is thus given by the Fourier-transform of the wave
function $\phi _{0}(\mathbf{r})$. The resulting momentum
distribution is then added to the momentum profile of the normal gas
which is typically at the edge of the trap. We plot the results of
our calculation in Fig. 2 (b)(d)(f). In general, the peaks are now
sharper compared to the case of an ideal Bose gas, which is just as
expected. A bimodal distribution still shows up when the condensate
fraction becomes non-zero (Fig. 2(d)). In real experiments, due to
the finite resolution in imaging, the height of the central
condnesate peak will be significantly reduced, and the bimodal
structure will become more pronounced and should be readily
observable.

After turnoff of the trap, the atomic momentum distribution will
be broadened during the time-of-flight process due to the
collision interaction. The interaction broadening of the momentum
distribution happens dominantly through the condensate fraction
\cite{2,4}, for which one has a higher number density and a
smaller expansion speed. To incorporate this effect, we use the
Gross-Pitaevskii equation \cite{9} to solve the evolution of the
condensate
wavefunction $\phi _{0}(\mathbf{r,}t)$, starting at $\phi _{0}(%
\mathbf{r})$ when the expansion time $t=0$. The Fourier component of
$\phi _{0}(\mathbf{r,}t)$ after a long enough expansion gives the
final momentum distribution for the condensate part. The results of
the calculation are shown in Fig.2 (b) (d) for $^{87}$Rb atoms. One
can see that the interaction broadening leads to some quantitative
corrections of the profile by lowering its peak value, but it does
not change much the overall picture. In particular, the bimodal
structure remains similar when there is a condensate fraction.

In summary, we have shown through explicit calculation that the
interference peaks, combined with the bimodal profile of the central
peak in the first Brillouin zone, provide an unambiguous signal for
the condensation phase transition for cold atoms in an optical
lattice at finite temperature. We develop some techniques to
calculate the atomic density profile, in particular for an
inhomogeneous system with a global harmonic trap. In general, the
density profile of the interference peak gives more detailed
information of the system, and a comparison of the density profiles
from the theoretical calculation and from the experimental
observation will contribute to the understanding of this strongly
correlated system at finite temperature.

This work was supported by the NSF awards (0431476), the ARDA under ARO
contracts, and the A. P. Sloan Fellowship.

\end{document}